# Interactive Overlays: A New Method for Generating Global Journal Maps from Web-of-Science Data

Loet Leydesdorff [a] & Ismael Rafols [b]

**Abstract**

Recent advances in methods and techniques enable us to develop an interactive overlay to the global map of science based on aggregated citation relations among the 9,162 journals contained in the *Science Citation Index* and *Social Science Citation Index* 2009 combined. The resulting mapping is provided by VOSViewer. We first discuss the pros and cons of the various options: cited *versus* citing, multidimensional scaling *versus* spring-embedded algorithms, VOSViewer *versus* Gephi, and the various clustering algorithms and similarity criteria. Our approach focuses on the positions of journals in the multidimensional space spanned by the aggregated journal-journal citations. A number of choices can be left to the user, but we provide default options reflecting our preferences. Some examples are also provided; for example, the potential of using this technique to assess the interdisciplinarity of organizations and/or document sets.

[a] Amsterdam School of Communications Research (ASCoR), University of Amsterdam, Kloveniersburgwal 48, 1012 CX Amsterdam, The Netherlands; loet@leydesdorff.net; http://www.leydesdorff.net .
[b] SPRU (Science and Technology Policy Research), University of Sussex, Freeman Centre, Falmer Brighton, East Sussex BN1 9QE, United Kingdom; i.rafols@sussex.ac.uk .

## 1. Introduction

In this study, we extend our previous construction of an interactive overlay which enables users to assess a document set—downloaded from the Web of Science (WoS)—in terms of its journal composition. In the previous overlay (Leydesdorff & Rafols, 2009; Rafols *et al*., 2010) we used the 222 ISI Subject Categories for this purpose. Although the maps using these categories can perhaps be considered reliable at a sufficiently aggregated level, in many cases; particularly, in the case of smaller sets, one would like to use a more detailed journal map instead. However, the *Journal Citation Reports* 2009 (JCR) of the (*Social*) *Science Citation Index* (SCI) contains 9,162 journals. Hitherto, the mapping of this comprehensive set and its clustering has been beset with problems (Leydesdorff, 2006). More recently, technical and methodological advances enable us to provide this interactive overlay.

The global mapping of the sciences using aggregated journal-journal citations has been on the research agenda of bibliometrics since the mid-eighties (Doreian, 1986; Doreian & Fararo, 1985; Leydesdorff, 1986 and 1987; Tijssen *et al*., 1987; cf. Price, 1965). At the time, several research teams concurrently realized that the data aggregated in the JCR make it possible in principle to generate maps of science at the journal level, as distinct from the maps of documents that had been developed as co-citation (e.g., Small & Garfield, 1985) or co-word mappings (Callon *et al*., 1983 and 1986) in previous years. In this experimental phase, however, maps were technically limited to local maps, for example, at the disciplinary or specialty level.

Journals as intellectual forms of organization of the scientific archive were first recognized by pioneers of bibliometrics such as Derek de Solla Price (1965), Francis Narin (e.g., Narin *et al*., 1972), Tibor Braun (e.g., Zsindely *et al*., 1982), and Eugene Garfield (e.g., 1972). Most scholars of this first generation focused on journal standing, hierarchy, and impact. Doreian (1986) developed his first journal maps from this perspective. In addition to "being cited," however, the JCR also provides access to the aggregated citing behavior and thus the performative power of journals in organizing the sciences intellectually (Latour, 1987).

The clusters and consequent configurations in journal maps can show how journal groups develop as intellectual organizers in niches (Leydesdorff & Van den Besselaar, 1998; Lucio-Arias & Leydesdorff, 2009; Van den Besselaar & Leydesdorff, 1996). These niches can robustly—that is, relatively independently of the precise clustering algorithms—be recognized as specialties. Neighbouring specialties are often weakly connected to one another across intellectual borderlines. Specialties develop concurrently and therefore heterarchically—that is, with different types of overlap and without an ordered nesting of categories. The exceptions are "interdisciplinary" journals which may relate across disciplinary borders at a next-higher level—with multidisciplinary journals such as *Nature* and *Science* at the top (Carpenter *et al*., 1973; Leydesdorff & Bornmann, in preparation)—or at the bottom with reference to engineering, the clinic, or other practical applications (Gibbons *et al*., 1994; Leydesdorff & Jin, 2005). The emergence of new journals can also serve as an indicator of new developments in science by means of "visual analytics" (e.g., Cook & Thomas, 2005; Thomas & Cook, 2006; cf. Leydesdorff & Schank, 2008). Such tools can be used as indicators for policy purposes in the early stages of an emergent field and/or technology (Leydesdorff *et al*., 1994).

Developments at the journal level are largely beyond the control of individual authors or funding agencies. Journals also operate in a market that is populated with publishing houses, libraries, etc. Authors can contribute to and change the reproduction of these structures by publishing new manuscripts. However, the dynamics of the communications aggregated at the level of journals are non-linear. Aggregated journal citation patterns among major journals, for example, are often stable over the years. In other words, they develop most of the time incrementally and recursively, building on previous citation patterns. The development of specialties (in terms of sets of journals) can be expected to reinforce borderlines by signaling the emergence of a symbolic order in the respective discourses (e.g., Leydesdorff & Probst, 2009; Rice, 1990).

Perhaps, in this age of the internet and Google Scholar where authors can read across journals more easily, this organizing power of journals may have waned (Bollen *et al*., 2005), but counteracting this is an increased emphasis in research management on legitimacy and citation impact. Institutional incentives point towards publishing in journals with the highest possible prestige (Halffman & Leydesdorff, 2010). Prestige is often associated first with inclusion in the (*Social*) *Science Citation Index,* or to a lesser extent in Scopus (because the latter database covers a considerably larger—and therefore less restricted—number of journals). The internet interfaces of these databases, and the freely available program *Publish or Perish*,[1] respectively, allow researchers to track their publication and citation records in these various domains.

In the 1980s and the early 1990s attempts to map the sciences were limited by the graphic interfaces of the line-based DOS and Unix environments, and by the intricacies of plotting alternatively by using mainframe computers. The versions of UCINet available at the time, for example, included multidimensional scaling (MDS), but the plotting power was confined to the 26 upper-case and 26 lower-case character sets for the labeling of 52 data points. The advent of the new technologies from Apple and Windows (in 1995) brought more ambitious programming efforts, such as the development of the network analysis and visualization program Pajek in 1996.[2] This higher resolution and increased computer power triggered a new spurt in developments during the first decade of the 2000s.

Boyack *et al.* (2005), for example, used large-scale computational power (VxOrd) to organize journals for the first time into a global map which these authors called programmatically "the backbone of science." Rosvall & Bergstrom (2008) developed a global journal map based on statistical decomposition analysis (Leydesdorff, 1991 and 2004; Theil, 1972). Rosvall & Bergstrom (2010) extended this approach to include the dynamic analysis. Still, the results of the algorithmic clustering remained unconvincing (Rafols & Leydesdorff, 2009), while the choice of similarity criteria has remained a cause for concern (Ahlgren *et al*., 2003). Although at a large scale the global maps may seem robust (Klavans & Boyack, 2009), the lower-level structures (for example, at the journal level) can depend on small changes in the relevant parameters (Klavans & Boyack, 2006; Leydesdorff, 2006 and 2008).

[1] Publish or Perish is available at http://www.harzing.com/pop.htm.
[2] Pajek is available for non-commercial usage at http://vlado.fmf.uni-lj.si/pub/networks/pajek/.

The mapping results can be visually impressive (e.g., Börner, 2010), but the analytical implications of parameter choices are not always under control: visualization necessarily involves the reduction of a multidimensional space into the two dimensions of a sheet or screen. Community-finding algorithms based on modularity (Blondel *et al*., 2008; Newman, 2006) may enable us to decompose large matrices in terms of their nearly-decomposable components, but the results depend also on a randomly chosen seed number, and therefore cannot be reproduced unambiguously. Are the fluctuations among the solutions indicative of relevant changes in the data, or are they statistical effects? The solutions may also remain dependent on the parameter choices made by the users (Van Eck *et al*., 2010).

Two solutions to the problem of the projection to a two-dimensional map have prevailed: multidimensional scaling (MDS) based on stress-minimization (Kruskall, 1964; Kruskal & Wish, 1978) or minimization of an energy function in the graph conceptualized as a system of springs (Kamada & Kawai, 1989; Fruchterman & Reingold, 1991). MDS can be considered as equivalent to factor analysis or principal component analysis (Schiffman *et al*., 1981, at p. 13), but the objective is visualization. Some network visualization programs (such as UCINet and Visone) offer MDS directly as a possible subroutine for the visualization.

Using MDS, the network is first conceptualized as a multi-dimensional space that is then reduced stepwise to lower dimensionality. At each step, the stress increases using, for example, Kruskall's stress function formulated as follows:

$$S = \sqrt{\frac{\sum_{i \neq j} (\|x_i - x_j\| - d_{ij})^2}{\sum_{i \neq j} d_{ij}^2}} \quad (1)$$

In this formula $\|x_i - x_j\|$ is equal to the distance on the map, while the distance measure $d_{ij}$ can be, for example, the Euclidean distance in the data under study (Borgatti, 1997). The choice of a distance or similarity measure remains open. In the 1980s and 90s, for example, one of us used MDS to illustrate factor-analytic results in tables, and in this case the Pearson correlation obviously provided the best match (e.g., Leydesdorff, 1986).

Spring-embedded or force-based algorithms can be considered as a generalization of MDS, but were inspired by further developments in graph theory during the 1980s. Kamada and Kawai (1989) were the first to reformulate the problem of achieving target distances in a network in terms of energy optimization. They formulated the ensuing stress in the graphical representation as follows:

$$S = \sum_{i \neq j} s_{ij} \text{ with } s_{ij} = \frac{1}{d_{ij}^2}(\|x_i - x_j\| - d_{ij})^2 \quad (2)$$

Equation 2 differs from Equation 1 by taking the square root in Equation 1, and because of the weighing of *each* term with $1/d_{ij}^2$ in the numerator of Equation 2. This weight is crucial for the quality of the layout, but defies normalization with $\sum d_{ij}^2$ in the denominator; hence the comparability between the two stress values.

The ensuing difference at the conceptual level is that spring-embedding is a graph-theoretical concept developed for the topology of a network. The weighing is achieved for each individual link. MDS operates on the multivariate space as a system, and hence refers to a different topology. In the multivariate space, two points can be close to each other without entertaining a relationship (Burt, 1982; Granovetter, 1973). For example, they can be close or distanced in terms of the correlation between their *patterns* of relationships (Burt, 1995).

In the network topology, Euclidean distances and geodesics (shortest distances) are conceptually more meaningful than correlation-based measures. In the vector space, correlation analysis (factor analysis, etc.) is appropriate for analyzing the main dimensions of a system. The cosines of the angles among the vectors, for example, build on the notion of a multi-dimensional space. The remaining discussion is over the choice of using Pearson correlations or cosine values (Egghe & Leydesdorff, 2009). In bibliometrics, Ahlgren *et al*. (2003) have argued convincingly in favor of the latter because of the skewedness of the citation distributions and the abundant zeros in citation matrices.

Technically, one can also input a cosine-normalized matrix into a spring-embedded algorithm. The value of (1 – cosine) is then considered as a distance, albeit in the vector space (Leydesdorff & Rafols, 2011a). In sum, there is a wealth of possible combinations in the parameter space of clustering algorithms and similarity criteria. Analytical arguments and not the esthetics of the representation have to guide us toward the choices to be made.

## 2. Methods and data

The data for this project was harvested from the Internet version of the *Journal Citation Reports* 2009 in August 2010. On the basis of this data an aggregated journal-journal citation matrix of 9,162 journals was constructed. Of the $9{,}162^2$ = 83,942,244 cells only 2,790,876 (= 3.32%) are filled with values larger than zero; the grand total of the matrix is 33,190,645 citations, or on average 11.89 per cell with a value larger than zero. This data was gathered from the "citing" side. Although the long tails of low values are sometimes summed on this (citing) side as "all others," the file contains 1,209 cells with a value of one. The cutoff at the lower end varies in the JCR with the sizes of the tails.[3]

The matrix was transformed into a cosine-normalized matrix both in the being-cited and the citing dimensions. Matrices with thresholds for the cosine values can then be exported in formats that can be read by the various visualization programs. We used SPSS for the cosine-normalization and Pajek/UCINet for the data manipulation. VOSViewer (based on an MDS-like algorithm)[4] and Gephi (containing a spring-embedder)[5] will be compared below for the visualization.

[3] As of 2010, Thomson Reuters no longer provides the CD-Rom versions of the JCR which allowed for more precise database management.

For example, after normalization in terms of the "being cited" patterns and with a threshold of cosine > 0.2, the matrix contains 346,771 (off-diagonal) values larger than zero. More than 99% (346,746) of these edges are contained within a largest component of 8,817 nodes (96.23%). The other nodes are mainly isolates. At the threshold level of cosine > 0.5, however, only 5,954 nodes (64.99%) are connected by 51,030 edges (less than 15%) in the largest component. We use this latter (much smaller) matrix to first sort out our methodological questions—e.g., about the choice of the visualization software—and then return to the larger set for the overlay map. Visualization software uses largest components because the isolates and non-related components cannot be positioned unambiguously with reference to the larger set.

As noted, we focus first on two programs recently made available for the visualization: VOSViewer (Van Eck & Waltman, 2010) and Gephi. Both programs are Java-based and thus their capacity is limited only by the hardware configuration. We worked with a standard desktop PC with 4 GB internal memory under Windows 7, 32-bits. This PC has problems handling the large component of 8,817 journals at the threshold level of cosine > 0.2, although Gephi manages to handle the large file after a long processing time. VOSViewer gave no error message for this file size using 8 GB and a 64-bits machine.[6]

[4] VOSViewer is freely available at http://www.vosviewer.com.
[5] Gephi is freely available at http://gephi.org.
[6] Ludo Waltman at CWTS (in Leiden) was so kind as to run our files at a large machine with 64 GB. The results were similar (*personal communication*, 26 April 2011).

One needs to generate these results only once since they can be saved as the coordinates of the basemap (to be discussed below).

Following up on the idea of an interactive global science map using the ISI Subject Categories (Leydesdorff & Rafols, 2009; Rafols *et al*., 2010), our objective in this study is to generate an equivalently interactive global map at the journal level. The aggregation of journals to the ISI Subject Categories is beset with error (Boyack *et al*., 2005; Pudovkin & Garfield, 2002). Rafols & Leydesdorff (2009) have shown that different classifications lead to globally similar maps, and suggested that the error is therefore stochastic and not so relevant at the aggregated level. However, for mapping against a baseline one may wish to use relatively small or specifically defined sets, and in such cases error matters.

There are cases in which one needs more finely-grained, journal-based maps instead of maps based on disciplines. If the publication subset to be positioned is relatively small (e.g., below hundreds) and/or dispersed, in such cases the lack of accuracy in the delineation of SCs matters. The journal level would be not only more informative, but also more precise. A second case is when mapping interdisciplinary fields, where the assigment of journals to SCs is sometimes controversial (e.g., nanoscience and nanotechnology). Here, a journal map may provide a different perspective on the fields under study, one not necessarily convergent with that of the broader classifications.

As with our previous toolkit, the basemap should provide the option of using any set downloaded from the Web of Science (WoS) and visualize the sets esthetically in terms of a global map of science. It would be best to have this option available interactively at the Internet. Gephi has an option to export files in the so-called gexf format which can be embedded in the html using GEXFExplorer. Leydesdorff *et al*. (2011) used this option to visualize the aggregated journal citation structures of the *Arts & Humanities Citation Index* 2008 at the Internet. Furthermore, a spring-embedder—the algorithm of Fruchterman & Reingold (1991)—is available in Gephi, and the algorithm for modularization of Blondel *et al*. (2008) is available for clustering the data; the resulting clusters can be colored accordingly. In summary, the graphic options of Gephi are state-of-the-art. As noted, the program is also efficient in handling large-size files.

The "VOS" in VOSViewer stands for "visualization of similarities." The algorithm used for this is akin to that of MDS: VOSViewer minimizes a stress function at the systems level (Van Eck *et al*., 2010). More recently, Van Eck & Waltman (2010) have integrated a clustering algorithm into the program that operates on the basis of the same principles as the positioning of the nodes. The cluster results are automatically colored into the map. The clustering algorithm operates with a parameter (γ) that can be changed interactively in order to generate more or fewer clusters in the solution. Additionally, a representation of the map as a density or heat map is provided in VOSViewer.

From our perspective, VOSViewer is closer to our objectives than Gephi because it relies on an MDS-like algorithm to position the data in the multidimensional vector space, whereas Gephi provides graph-analytical routines that operate on the networked relations. However, both programs offer a wealth of possibilities for the visualization of a comprehensive map. We use the (smaller) set—normalized with cosine > 0.5—in the cited dimension first for comparison of the options, then provide arguments for our choice of VOSViewer, and change thereafter to the larger set of 8,800+ journals included in the largest component when the threshold is set at cosine > 0.2.

## 3. Construction of the basemap in VOSViewer or Gephi?

### *3.1. VOSViewer*

VOSViewer allowed us to import the Pajek-file for the set of 5,954 journals without a problem. The larger set of 8,817 journals generated an error message because of memory shortage, as did also the in-between sets for cosine > 0.3 (8,378 journals) or cosine > 0.4 (7,504 journals). Extending the available memory by changing the parameter for the Java .jar file with 8 GB and 64-bits solved this problem. Thus, it appears that the generation of the basemap will soon be within the reach of an average user of VOSViewer.

The current limitation is a minor disadvantage because the coordinates of the global map can be made available as an ASCII text file that can be loaded into the program or webstarted at the Internet. Another minor disadvantage of VOSViewer is perhaps that the computation using MDS-type positioning requires a longer time than using spring-embedded algorithms.

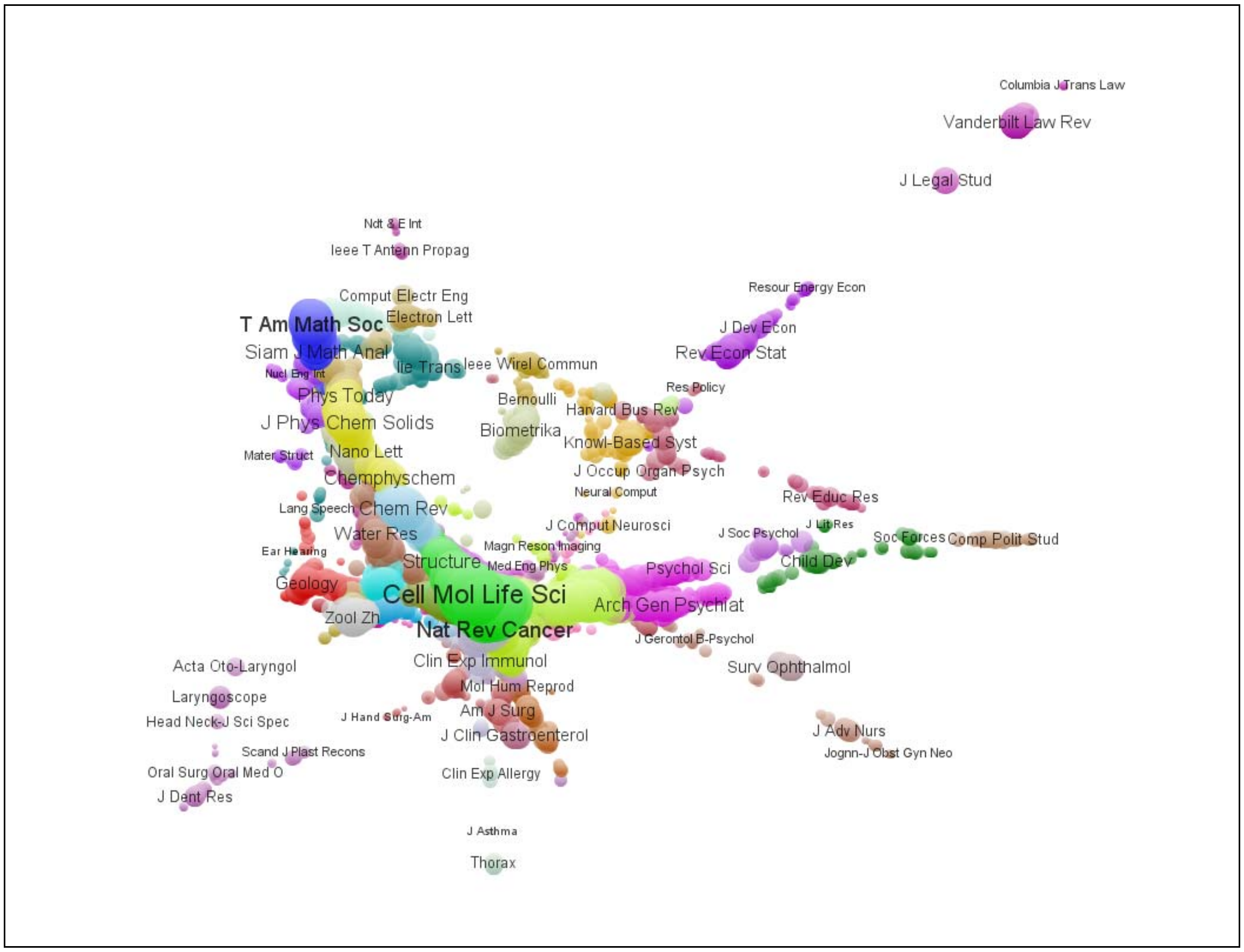

**Figure 1**: 5,954 journals similar in their being-cited patterns above cosine > 0.5; 70 colors (clusters) distinguished by VOSViewer (γ = 1).

Figure 1 provides the basemap for the file of 5,954 journals that are similar in their being-cited patterns above cosine > 0.5. In addition to the torus-like construct normally found in these maps (Klavans & Boyack, 2009), this visualization also suggests axes. At the top right, for example, some law journals are set apart. The bio-medical cluster at the bottom left shows a fine structure. Seventy clusters are distinguished and colored differently.

One can associate the right upper half of the figure with the social sciences and the left-lower half with the natural and life sciences, which are then aligned on a continuum from mathematics at the top to specialties in clinical medicine at the bottom. However, the distinctions among these larger structures may be caused by the relatively high value of the threshold (cosine > 0.5) which partitioned the network into 70 communities. The unfolding algorithm of Blondel *et al*. (2008) distinguished more than 400 communities ($Q \approx 0.83$) in this same data. As noted, we shall use a lower threshold value for the overlay below.

*3.2 Gephi and GEXFExplorer*

Although Gephi had no problems importing the larger file of 8,817 journals (cosine > 0.2), the exported file in gexf becomes so large (> 60 MB) that using this file at the Internet may lead to problems. Thus, we worked again with the smaller file of 5,954 journals (cosine > 0.5). Because of this high threshold, the number of communities detected by the modularity algorithm (Blondel *et al*., 2008) increases to slightly above 400, with modularity values ($Q$) between 0.82 and 0.84. Using the larger file (cosine > 0.2), the connections are denser and modularity drops to values of $Q \approx 0.64$ distinguishing approximately 160 communities.

As noted, these values are statistical. Consequently, one cannot obtain twice the same output—and hence coloring—using this algorithm. In a previous project (Rafols &

Leydesdorff, 2009), we found it impossible to make comparisons across years using this algorithm. The community-finding algorithm not oly generates fluctuations (because of using random seeds), but the data may also fluctuate from year to year (Renaud Lambiotte, *personal communication*, 16 December 2009).[7]

Gephi comes with two algorithms for generating the visualization that are particularly suited to our purpose: the spring-embedder of Fruchterman & Reingold (1991) and a new algorithm called ForceAtlas that was developed by the Gephi team for this purpose. Sébastien Heyman of the Gephi team was so kind as to run our file on his computer using the second version of ForceAtlas currently still under development (*personal communication*, 23 April 2011). The result is impressive in terms of the dissolution into the 400+ communities. Figure 2 provides a strong enlargement (800%) of the environment of information science journals.

[7] Rosvall & Bergstrom (2010) claimed that nephrology, although separate in 2001, joined the area of medicine in 2003, but more or less departed from it again in 2005. Because this was tested for significance in each year, the authors claim that these changes are significant. Without significance testing, however, the visible changes may be due to fluctuations in the intensity with which authors in nephrology cite and are cited in the literature in general medicine. Such problems may be unavoidable in the dynamic map.

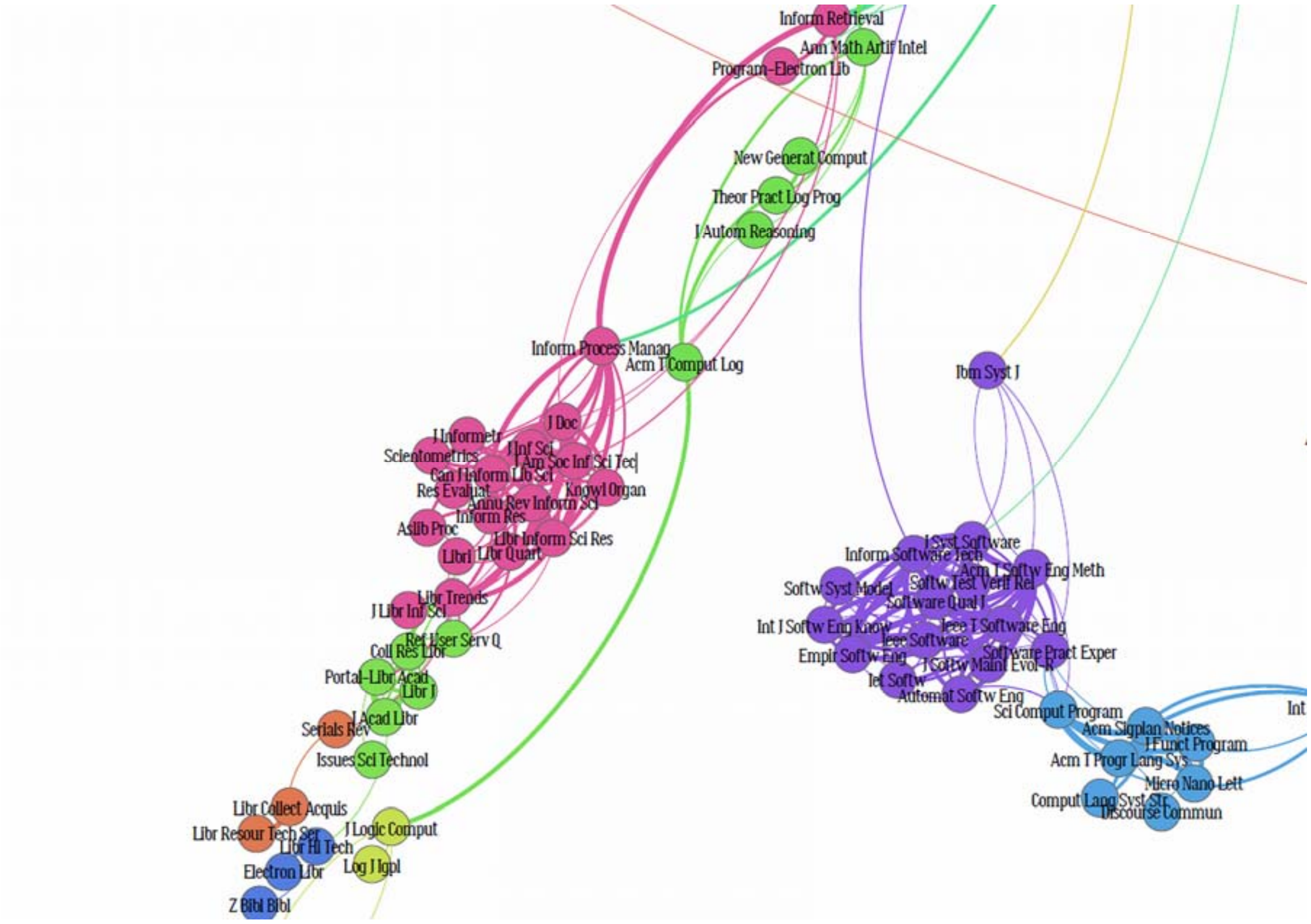

**Figure 2**: Details of the map of 400+ communities (modularity 0.83); enlarged 8 times.

Two problems remained:

1. The size of the map in relation to the size of the labels: the labels in the enlargement in Figure 2 are far too small to read in the overall map, and the labels become cluttered at a larger scale. The global map is available online at http://www.leydesdorff.net/journalmaps/forceatlas.htm. At the bottom left one can adjust the node size, but unfortunately within GEXFExplorer the trade-off between label size and readability has not yet been solved (Alexis Jacomy, *personal communication*, 31 January 2011). By zooming in, however, one obtains a better representation. Alternatively, one can download the gexf file (available at http://www.leydesdorff.net/journalmaps/forceatlas.gexf) and read this file into Gephi, which is freely available at http://gephi.org. Thus, one can obtain the full power of Gephi to change parameters in the visualization (such as node and label sizes, clustering algorithms, etc.).

2. ForceAtlas focuses on sorting the communities apart, while appreciating intergroup linkages, but these tend to become very long edges because of their separation. Using Fruchterman & Reingold's spring-embedder instead within Gephi provides a more compact representation (available at http://www.leydesdorff.net/journalsmaps/fruchreing.htm, and analogously as a gexf file at http://www.leydesdorff.net/journalmaps/fruchreing.gexf); but in this case the substantive separation among the communities remains sometimes confused. The problem of the relative size of the labeling becomes even greater when using this solution.

In summary, it is not possible to distinguish clearly among labels that are close to one another on overlays using either of these layouts for these large sets. The two layouts require a choice between distinguishing more sharply among subsets (using ForceAtlas) or exhibiting more relations across components (using Fruchterman-Reingold). A lower threshold of cosine > 0.2 may make the problem of relations across sets less acute, but the problem of label overlap versus readability in terms of size will be further aggravated because instead of 400+ only about 160 communities are then distinguished.

## 4. The Cited Map as a baseline

Figure 3 shows the solution for 8,817 journals in terms of similarity of being cited within this set using the lower threshold level of cosine > 0.2 (cf. Egghe & Leydesdorff, 2009). This map is comparable to the map based on co-citations among 5,000 major journals provided by Van Eck & Waltman (2010, at p. 535), except that we rotated the map 180 degrees in order to keep our results comparable to the ones based on the ISI Subject Categories (Leydesdorff & Rafols, 2009; Rafols *et al.*, 2010). As in the latter case, this figure shows a torus-like structure (Klavans & Boyack, 2009).

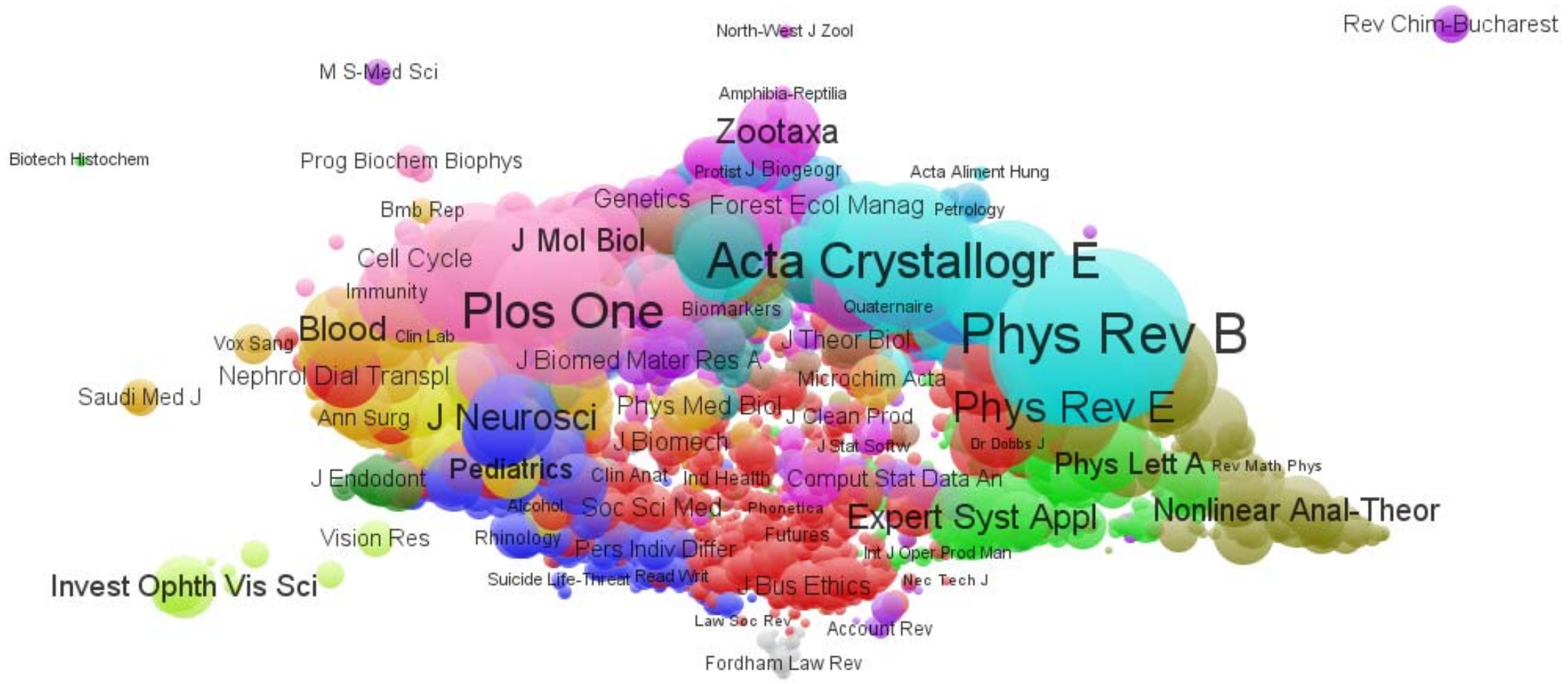

**Figure 3**: 8,817 journals similar in their being-cited patterns above cosine > 0.2; 21 clusters ($\gamma = 1$) using the algorithm built into VOSViewer (Van Eck *et al.*, 2010).[8] (Sizes of the nodes correspond to the respective numbers of publications in 2009.)

[8] One can webstart this map at http://www.vosviewer.com/vosviewer.php?map=http://www.leydesdorff.net/journalmaps/cited02.txt

The two large domains of the bio-medical and the natural sciences are connected in Figure 3 by two bridges: the biological and environmental sciences at the top and the social sciences at the bottom. Some computer-science and statistics journals are placed in the center, but the mathematics group is set apart to the right. The noted group of law journals is visible as a small gray-colored cluster at the bottom.

Only 21 communities are distinguished by VOSViewer at this level of connectedness (using $\gamma = 1$; as against 70 at the level of cosine > 0.5).[9] In other words, the system is more tightly knit than above (in Figure 1). However, this strong reduction in the number of colors (clusters) can be convenient for our purpose: our objective is to map at the level of journals and not in terms of (hundreds of) specialties. We shall discuss the quality of the clustering below in the case of the citing map.

Using the property that the coordinates of all nodes can be saved in a (comma-separated) database file, the abbreviations of the titles could be used to generate an interface between any download from the WoS and this map. Unfortunately, the abbreviations are not completely standardized between the JCR and the WoS. Thus, for the matching we use the full titles of the journals instead.

## 5. The Citing Map as a baseline

The same 9,162 journals can be evaluated alternatively in terms of their citing patterns. Whereas the "being-cited" patterns refer to the archive of science, "citing" is the running variable which is exclusively set to the current publication year. Journals are perceived in their environment mainly in terms of how they are cited as an archive (Bensman, 2007). The citing variable, however, does not accumulate.

[9] At $\gamma = 1.5$ and $\gamma = 2$, the number of clusters are 31 and 39, respectively.

The positioning in this (citing) dimension is coupled to the activity of the citing authors, and thus can indicate changes in citation patterns without a lag or dampening. For this reason of continuous updating by the community (of authors) itself, we preferred this direction for the map in the case of the previous overlay in terms of ISI Subject Categories. Citing represents the knowledge base of new knowledge claims, while cited represents the accumulated impact of journals on audiences. However, this choice can also be left to users according to their specific research questions.

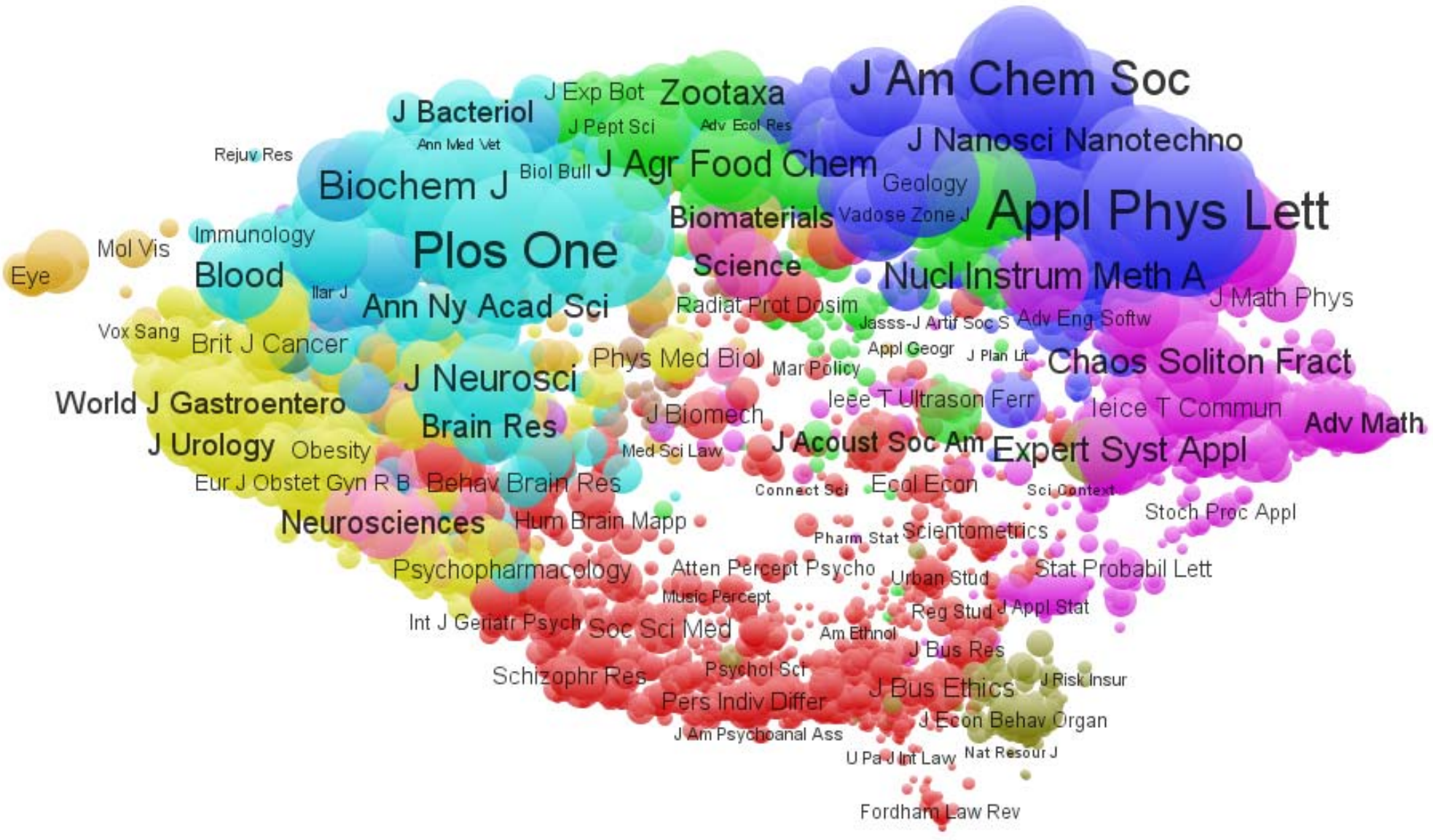

**Figure 4**: 8,860 journals similar in their citing patterns above cosine > 0.2;
19 clusters ($\gamma$ = 1) using the algorithm built into VOSViewer (Van Eck *et al.*, 2010).[10]
(Sizes of the nodes correspond to numbers of publications in 2009.)[11]

---

[10] At $\gamma$ = 1.5 and $\gamma$ = 2, the number of clusters are 77 and 121, respectively.
[11] One can webstart this map at
http://www.vosviewer.com/vosviewer.php?map=http://www.leydesdorff.net/journalmaps/citing02.txt

The largest "citing" component connects 8,860 of the 9,162 journals and contains 928,878 of the 928,882 edges (> 99.9%). In other words, the set is virtually complete. The network is less decomposable than the previous one, with a modularity of Q ≈ 0.475 distinguishing approximately 40 communities (using Blondel *et al.*'s (2008) algorithm for the unfolding). In other words, in their citing behavior authors reach across specialty boundaries, but this can be considered as variation, while the accumulated structures in the cited journals are reproduced more regularly as boundaries in the intellectual organization of the sciences.

| | *Clusters 1-19* | *Nr of journals* | *Factor 1-18 (Rafols et al. 2010)* |
|---|---|---|---|
| 1 | Social Sciences | 1962 | Biomedical Sciences |
| 2 | Agricultural Sciences | 1349 | Materials Sci. |
| 3 | Chemistry | 1342 | Computer Sciences |
| 4 | Clinical Medicine | 1177 | Clinical Medicine |
| 5 | Computer Sciences | 1062 | Economics, Politics and Geography |
| 6 | Biomedical Sciences | 864 | Psychology |
| 7 | Infectious Diseases | 288 | Ecological Sci. |
| 8 | Biophysics | 285 | Chemistry |
| 9 | Economics & Management Sci. | 266 | Geosciences |
| 10 | Neurosciences | 173 | Cognitive Sciences |
| 11 | Occupational Hygiene | 34 | Health and Social Issues |
| 12 | Rheumatology | 33 | Engineering |
| 13 | Multidisciplinary | 16 | Environmental S&T |
| 14 | Biomedical Engineering | 4 | Agricultural Sci. |
| 15 | *Micro & Nano Letters* | 1 | Infectious Diseases |
| 16 | *Journals of Gerontology Series A-Biological Sciences And Medical Sciences* | 1 | Social Studies |
| 17 | *Pteridines* | 1 | Physics |
| 18 | *Pediatric Research* | 1 | Business and Management |
| 19 | *Turkish Journal of Biochemistry-Turk Biyokimya Dergisi* | 1 | |

**Table 1**: 19 clusters of journals in citing patterns using VOSViewer ($\gamma = 1$).

In Table 1 the 19 clusters distinguished by the algorithm in VOSViewer are compared with the 18 factors that we distinguished on the basis of factor analysis for coloring the overlay among the 222 ISI Subject Categories in Rafols *et al*. (2010). Five clusters contain only a single journal, and one has only four journals. These six clusters are barely visible in the map. Actually, only the first ten clusters can be considered meaningful, and these dominate the map. Using these colors, in our opinion, is informative, although this clustering algorithm is as yet far from convincing.

In a previous study, we discussed various possible classification systems of journals, yet without a conclusive recommendation (except for a note of caution against the use of examiner-based classifications such as the SCs; Rafols & Leydesdorff, 2009). One advantage of these new maps is that the user can change the classification system to suit specific needs. After reading the file citing02.txt (available at http://www.leydesdorff.net/journalmaps/citing02.txt) into Excel, the user can replace the column labeled "cluster" with another classification. For example, the file at http://www.leydesdorff.net/journalmaps/citing/blondel.txt contains the same mapping information, but uses the 41 communities distinguished by the algorithm of Blondel *et al*. (2008). This provided us with the following figure (Figure 5).

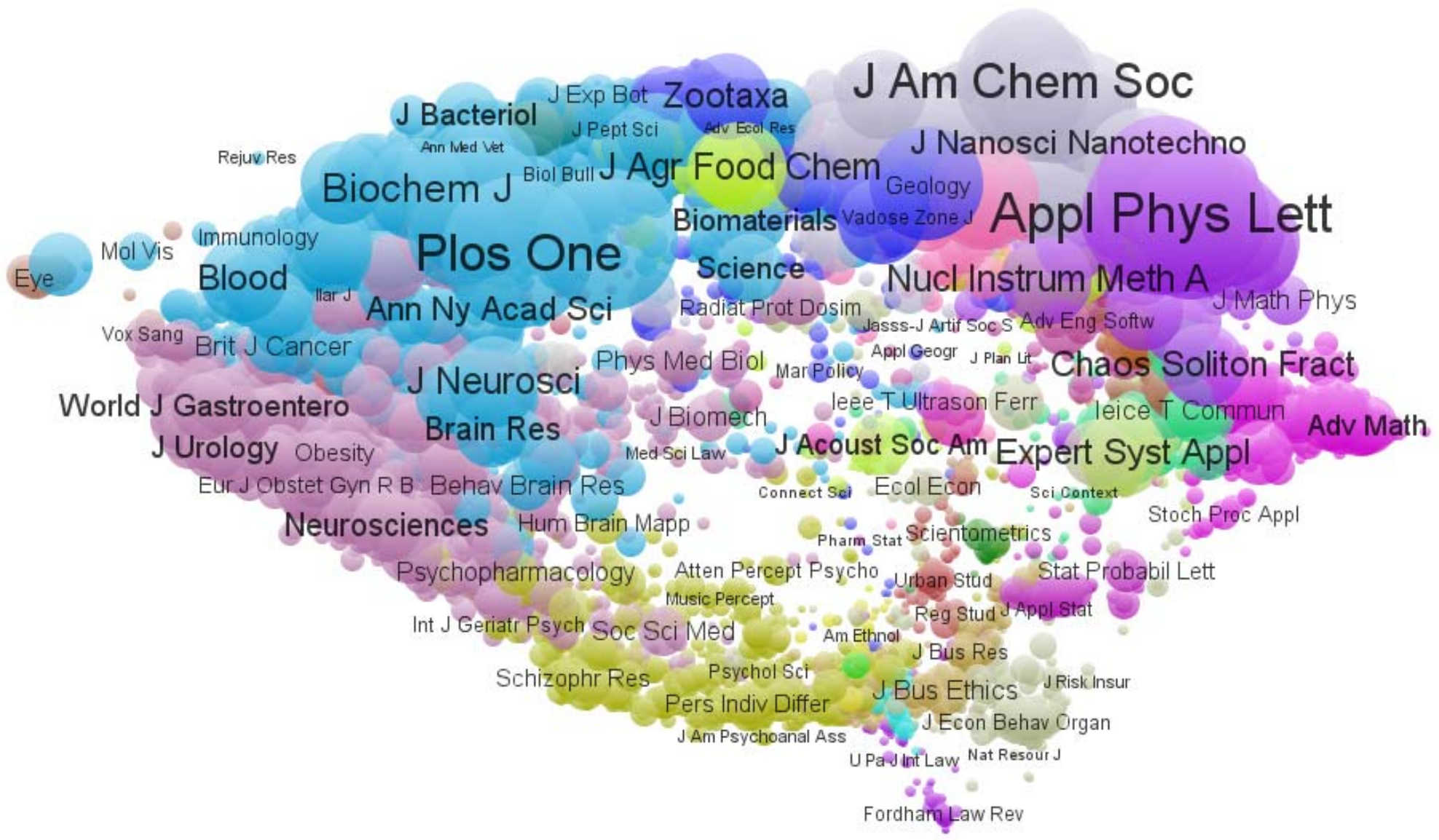

**Figure 5**: 8,860 journals similar in their citing patterns above cosine > 0.2; 41 clusters (colors) distinguished by the rapidly unfolding algorithm of Blondel *et al*. (2008). (Colors adapted within VOSViewer.)

The community structure is broadly similar, but more finely-grained particularly in the relatively poorly populated area contained by the torus. In summary, one can freely change the clustering and accordingly the coloring while maintaining the basemap for the overlay. The mapping and clustering are thus uncoupled. One analytical advantage of such uncoupling is to forestall potential erroneous interpretations arising from either the reduction of dimensions in the mapping or the aggregation in the clustering.

If one wishes to replace the *default* coloring for generating the overlays, the column "cluster" in the table files at http://www.leydesdorff.net/journalmaps/citing.dbf and/or http://www.leydesdorff.net/journalmaps/cited.dbf, respectively, must be changed. The programs for generating the overlays will then use the newly inserted values.

## 5. The generation of the overlay files

Two programs are made available online at http://www.leydesdorff.net/journalmaps/citing.exe and http://www.leydesdorff.net/journalmaps/cited.exe, respectively, for processing a file containing downloaded data from the WoS. These .exe files also need the table file citing.dbf and/or cited.dbf, respectively, in the same folder. (The table files can be downloaded from the same page.) In addition to the coordinate information for the maps, the full titles of the journals as provided by the JCR are listed in these files. As noted, the titles of the journals are used as keys for the matching. (In the case of an unforeseen mismatch one is advised to adapt the title in the corresponding table file.)

When the programs and tables are brought into a single folder with the input file, which is downloaded from the WoS and renamed "data.txt," an output file can be generated. This file is called either "cited.txt" or "citing.txt" depending on the routine in question. These files can be used as input to VOSViewer and thus visualized. Figure 6, for example, shows the publication profile of the first author of this article using the "citing" map for the projection.

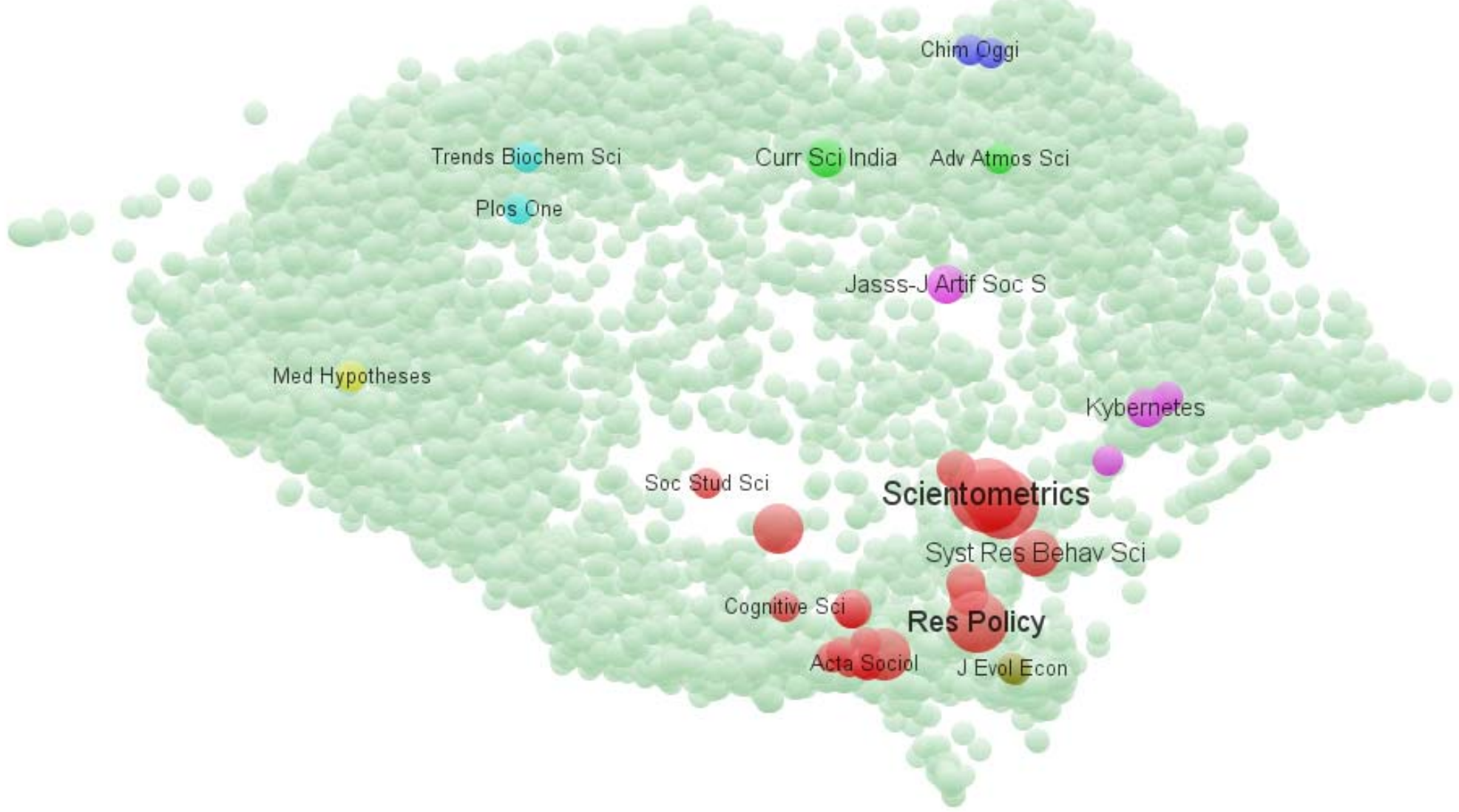

**Figure 6**: Overlay map of 168 articles published by the first author (Leydesdorff) in journals included in the ISI journal set.

The colors in the overlay map correspond to the colors in the global map provided above as Figure 4. Thereafter, all options available in VOSViewer are available to the user. Thus, in addition to this label view, one can choose for the density view or the scatter view, and the options enable one to make all labels equally visible, to change the colors of clusters, etc.

The default setting of our overlay programs presumes that the size of the nodes will be equal to the $\log_{10}(n + 1)$. The value of $n$ is augmented by one in order to prevent the disappearance of a node in the case of a single publication (since $\log(1) = 0$). Depending on the relative sizes, one may wish to use a function other than the logarithm. The values of $n$ can be found in the table file overlay.dbf which is generated at each run; this file can be read into Excel. By replacing the column labeled "weight" with the values in the column *NPubl* in the file overlay.dbf, one obtains a map with a linear relation between size and publication volume. However, one problem yet to be solved is that VOSviewer, as against Pajek, contains an algorithm for relative size normalization by dividing all weights by the average weight. This (currently obligatory) algorithm is problematic when comparing different overlays quantitatively (e.g., in an animation). For example, when analyzing the diffusion and growth of emergent topics, one cannot compare the visual maps, but only the underlying quantitative data (Kiss *et al*., 2010; Leydesdorff and Rafols, 2011b).

As noted, the file overlay.dbf also contains the Blondel classification. If one wishes to assume this classification (or any other one) as the default for generating overlays, if it is necessary to change the cluster indication in the tables cited.dbf and/or citing.dbf in this respect (in Excel or SPSS) and save these files thereafter as .dbf tables with the same name.

## 6. An Application: Mapping "Interdisciplinarity"

One of our motivations for developing these overlays has been the wish to evaluate interdisciplinary developments that are potentially relevant to science policy (Leydesdorff *et al*., 1994; Leydesdorff & Rafols, 2011a and b). Let us therefore test these newly developed overlay methods in two areas of contestation about the function of interdisciplinarity: (1) the assessment of interdisciplinary units such as university departments (Rafols *et al*., 2011), and (2) the delineation of newly emerging specialties such as nanoscience and nanotechnology (Leydesdorff & Schank, 2008).

### *6.1 Innovation Studies versus Business & Management*

In a recent study, Rafols *et al.* (2011) compared the research portfolios of business and management (BM) schools with those of innovation studies departments (many of which are sometimes located in business and management schools) in the case of the UK. We found that mainstream BM schools such as the London Business School focus their publications and citations narrowly in the core journals of business and economics, while departments that are oriented towards innovation studies, publish across disciplinary lines (e.g., SPRU, the Science and Technology Policy Research Unit at the University of Sussex). This analysis was also performed using overlay maps based on the ISI SCs, and provided quantitative evidence that the evaluation tends to rank mono-disciplinary output more highly than interdisciplinary sets, as qualitative studies had previously suggested (Laudel & Origgi, 2006).

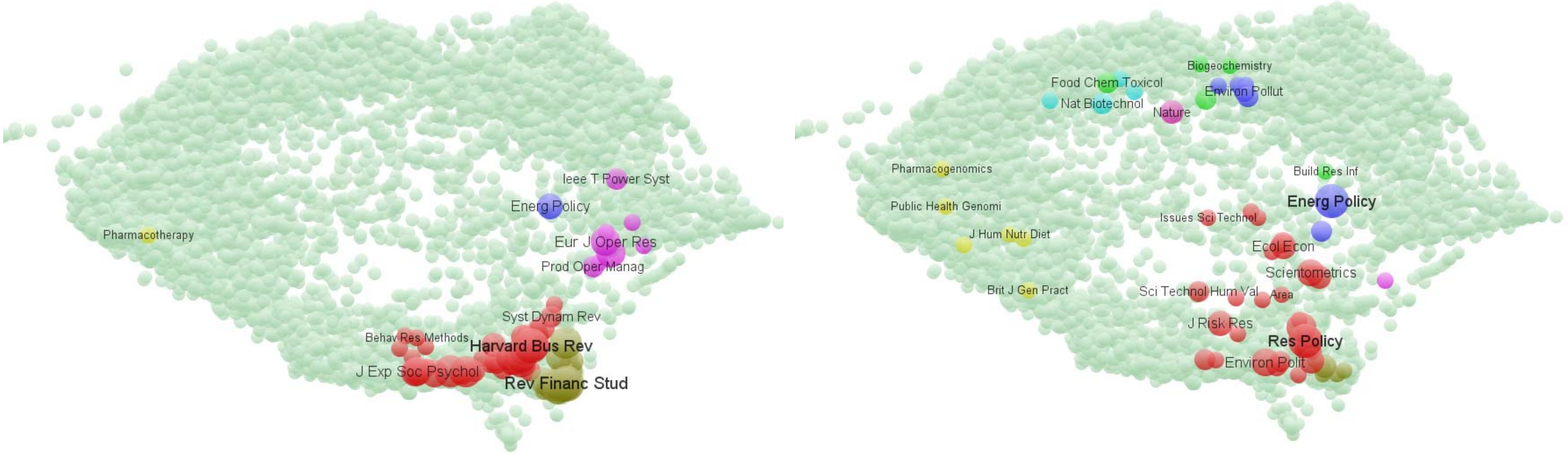

**Figure 7**: Overlay maps comparing journal publication portfolios from 2006 to 2010 between the London Business School (on the left) and the Science and Technology Policy Research Unit SPRU at the University of Sussex (on the right).

Figure 7 shows that SPRU publication (on the right side of Figure 7) cover a diverse range of disciplines, whereas the London Business School (to the left side) has a heavily concentrated profile that focuses to a much larger extent on business and management studies journals. These representations are "static." One can animate the representation by downloading data for a sequence of years, generate the map for each year, and save the screenshot into PowerPoint (Leydesdorff and Rafols, 2011b).

Figure 7 shows visually the comparison between these two organizations that could also be made in terms of numbers. However, a table or bar chart cannot show the interdisciplinarity of a set, in the sense that it does not convey information about whether the journals in questions are proximate or distant in cognitive space (Rafols *et al.*, 2010; Leydesdorff & Rafols, 2011a and b). The visualization enables us to recognize at a glance differences in patterns with reference to the full journal set. In the case of journal maps, the quality of the representation is more precise and convincing than when using the SCs, since the journal names in the individual documents are used directly for the mapping. The so-called "indexer effect" is thus avoided (Rafols & Leydesdorff, 2009).

### *6.2 Nanoscience and nanotechnology*

After the emergence of nanoscience and nanotechnology at the turn of the century (e.g., Leydesdorff & Schank, 2008), the ISI of Thomson Reuters introduced a new subject category in 2005 that contained 27 journals. In 2009, 59 journals were subsumed under this category. How heterogeneous is this category in terms of disciplinary affiliations?

The new overlay tool enables us to position these journals in relation to larger journal structures.

For this purpose we first generated the aggregated journal-journal citation matrix among 58 of these 59 journals,[14] and factor-analyzed it. Four factors in the cited patterns explain 35.9% of the variance. We used these factor delineations to divide the journals into four groups: "chemistry and condensed matter physics" (20 journals with primary factor loadings), "microfluidic devices" (7 journals), "nanomedicine" (7 journals), and "electronics" (11 journals). Only journals with factor loadings larger than 0.4 on a specific factor were used to color the map in Figure 8, for the sake of clarity.

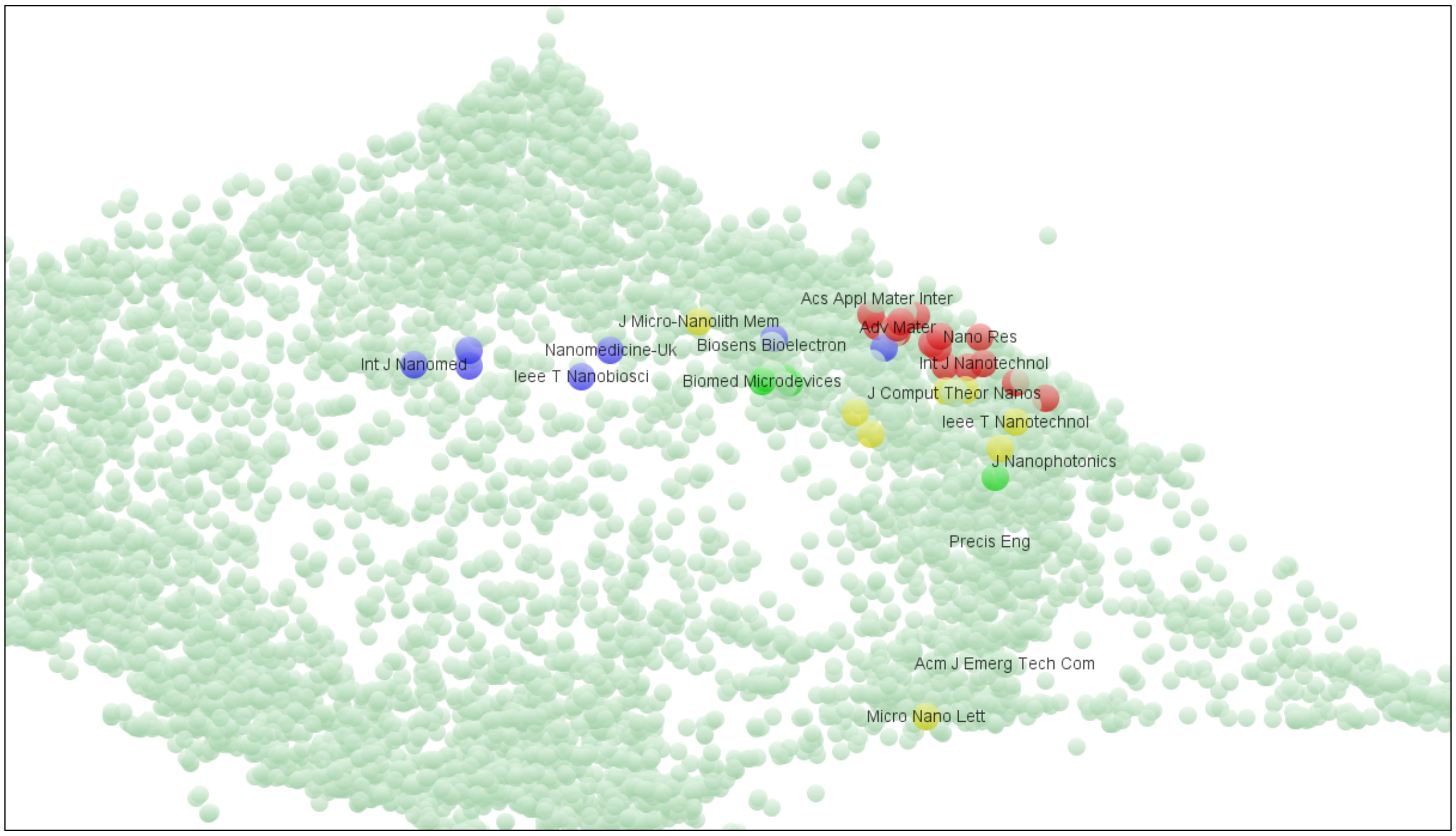

**Figure 8.** Main factors in the journal set of Nanoscience and Nanotechnology, overlaid on the cited map.

[14] *ACS Applied Materials & Interfaces* was added in 2009 and is not yet covered by the JCR 2009.

Figure 8 provides a visual answer to the question of whether and how closely these disciplinary-specific journals are related to the intellectual organization of the journal structures of nanoscience & nanotechnology: journals related to "nanomedicine" are colored purple (left); those related to "microfluidic devices" green (at the centre); those related to "chemistry and condensed matter physics" red (upper right); and those associated with "electronics" yellow (lower right). The large distances between the four groups raises questions about whether placing all these journals under a single "interdisciplinary" category is appropriate from the perspective of science dynamics.

## Conclusions and discussion

The results of this exploration exceeded our expectations: VOSViewer can serve as a platform for interactive journal overlays. One can either use the default values to size and color or else feed the tables with one's own preferences. For example, various classification schemes can be used for the journals and the nodes can be sized differently. One drawback (which perhaps can be repaired in a future version) concerns the scaling of weight relative to the average weight. It would be nice if the facilities for sizing the nodes—like those for example available in Pajek—were provided because the sizes of nodes can influence the visual comparisons, for example, in animations (Leydesdorff & Rafols, 2011b).

VOSViewer allows for uploading the results on the Internet; for example, one can webstart the basemap in the citing direction using the following hyperlink:

http://www.vosviewer.com/vosviewer.php?map=http://www.leydesdorff.net/journalmaps/citing02.txt. The direct transport in the form of text files allows for easy reproduction and for changes in the figures.

Our exploration of the options with Gephi—the other freeware program—were less successful. A number of hurdles could be specified which may be solved in future versions of this program. However, the use of a spring-embedder was also conceptually less attractive for our project because of the different topologies involved. VOSViewer follows the MDS tradition and considers the network as a multivariate system in which individual nodes (journals) are positioned (Van Eck *et al*., 2010).

In the case of a spring-embedder the focus is on individual relations, and the positions of the node follow as a result of the spanning of the network among the normalized links. The topology is that of the network graph. The latter perspective can perhaps be considered as an advantage in the case of a local map, because in this case the system sometimes cannot be specified conclusively.

In general, the relations in the graph indicate where information is communicated, but the positions in the multidimensional space enable us to specify the position of the communication from a systems perspective. Furthermore, the cosine is a spatial measure best suited to map the multidimensional space. In summary, we see a great potential and no major problems with using VOSViewer for the purpose of interactive overlays at the level of journals.

## Acknowledgement

We are grateful to Ludo Waltman for running VOSViewer with the basemap on a large computer facility. Sébastian Heyman was so kind as to run our basemap with ForceAtlas2 in Gephi. Andrea Scharnhorst, Nees-Jan van Eck, and Ludo Waltman provided comments on previous drafts.